\documentclass[twocolumn,showpacs,preprintnumbers,amsmath,amssymb,A4paper]{revtex4}

\usepackage{graphicx}% Include figure files
\usepackage{dcolumn}% Align table columns on decimal point
\usepackage{bm}% bold math

\begin{document}
\newcommand{\kp}{{\bf k$\cdot$p}\ }

\title{Spin-flip reflection of electrons from a potential barrier in heterostructures}
\author{Pawel Pfeffer and Wlodek Zawadzki}
 \affiliation{Institute of Physics, Polish Academy of Sciences\\
Al.Lotnikow 32/46, 02--668 Warsaw, Poland\footnotetext{ e-mail address: pfeff@ifpan.edu.pl}\\}
\date{\today}
\begin{abstract}
Spin-conserving and spin-flip opaque reflections of electrons from a potential barrier in heterostructures are described. An electric field of the barrier is considered to be the only source of energy spin splitting in the presence of spin-orbit interaction and its form is calculated in the three-level {\kp} model for a nontrival case of unbound electrons. Reflection angles and amplitudes are calculated for oblique incoming directions. It is shown that the system can serve as a source or filter of spin-polarized electron beams. Two unexpected possibilities are pointed out: a) non attenuated electron propagation in the barrier whose height exceeds the energies of incoming electrons, b) total reflection of electrons whose energies exceed barrier's height.
\end{abstract}
% PACS, the Physics and Astronomy
% Classification Scheme.
\maketitle
\section{\label{sec:level1}INTRODUCTION\protect\\ \lowercase{}}

It is well known that the spin splitting (SS) of energies in various quantum systems is intimately related to their symmetry. According to the Kramers theorem, the energy of an electron in a periodic system satisfies the equality: $E_{{\bf k}\uparrow} = E_{-{\bf k}\downarrow}$, where $\bf k$ is the wave vector and the arrows $\uparrow$ and $\downarrow$ signify spin-up and spin-down projections, respectively. If, in addition, a system is characterized by the
inversion symmetry, the two spin states are degenerate for any $\bf{k}$ value: $E_{{\bf k}\uparrow} = E_{{\bf k}\downarrow}$. In general, however, for a given ${\bf k}$ direction and value SS may occur even without an external magnetic field. It was shown by Dresselhaus [1] that in bulk semiconductors of the zinc blende structure, which have no inversion symmetry, the conduction bands of III-V compounds are characterized by an anisotropic spin splitting
proportional to $k^3$ for small $k$ values. This splitting was extensively investigated in bulk semiconductors. With the advancement of semiconductor quantum structures a new type of inversion asymmetry became possible, namely the structure inversion asymmetry (SIA). The SIA mechanism also leads to SS of electron energies. The SS related to SIA, often called "the Rashba splitting", is of interest due to possible applications, as it can be influenced by an external electric
bias. This interest is a part of the wider movement aimed to use properties of electron spin for technical use.

The subject of SS related to SIA has a controversial history, as reviewed by the present authors [2]. In a widely quoted paper, Bychkov and Rashba [3] wrote
down the following Hamiltonian for SS in heterostructures, in which the growth direction is parallel to a high symmetry axis
\begin{equation}
{\hat H} = \alpha ({\bm {\sigma}} \times {\bf k})\cdot {\bm{\nu}}\;\;.
\end{equation}
Here $\bm{\sigma}$ are the Pauli matrices, $\bf{k}$ is the 2D electron wave vector transverse to the growth direction, $\bm{\nu}$ is the unit vector in the growth direction, and $\alpha$ is a coefficient. In their paper, Bychkov and Rashba did not mention the inversion asymmetry, but it is recognized by now that the
coefficient $\alpha$ has a nonzero value only if the system is characterized by a structure inversion asymmetry along the growth direction. The Hamiltonian given by Eq. (1) resembles that of the spin-orbit interaction (SOI), but one should not confuse it with the SOI in which an electric field $\bf {\cal E}$ appears instead of $\bm{\nu}$. As follows from the description given in [2], a consensus on the theoretical treatment of the problem of SS due to SIA has been reached, see also [4-6]. Namely, it has been demonstrated that the SS of conduction energies in asymmetric quantum wells (QWs) is mostly related to asymmetric offsets of the conduction and valence bands, both due to different energy gaps and spin-orbit energies. In particular, the SS of a conduction band is not proportional to the electric field in this band because the average electric field in a bound state is zero if one neglects a difference of the effective masses in the well and barriers [7].

It was recognized a few years ago that the SOI can be used to manipulate electron spins in semiconductor heterostructures. In particular, it was proposed to fabricate spin filters by either driving electrons through inhomogeneous heterostructures with different strengths of the SOI [8, 9] or by reflecting 2D electrons from a lateral potential barrier in an asymmetric quantum well [10 - 15]. Recently, the present authors described spin-flip and spin-conserving electron reflections from a potential barrier in a vacuum using the Dirac equation for relativistic electrons [16].

In this paper we are concerned with the spin-flip reflection processes in heterostructures. It was exprimentally demonstrated by Chen et al [10, 12, 14] that, in the presence of the SOI, an opaque reflection from a potential barrier separates electrons undergoing spin-flip reflections from those experiencing spin-conserving reflections.  In their system, the electron spin splitting was caused by SIA of asymmetric QWs and the interaction with the barrier.
In our paper we describe the spin-flip and spin-conserving reflections introducing three new elements. First, we consider the simplest system in which \emph{the only} source of electron spin-splitting is its interaction with the barrier. This can be realized either by using a symmetric QW or by performing experiments without 2D heterostructures. Second, we actually calculate the spin splitting of unbound electron energies due to the interaction with the barrier, which was not attempted in previous works. Third, we point out quite unusual possibilities of non-attenuated electron propagation in the barrier whose height exceeds the energies of incoming electrons and the total reflection of electrons whose energies exceed barrier's height.

Our paper is organized as follows. In Section II we consider the spin splitting of electron energies due to the interaction with the barrier. Section III contains analysis of spin-conserving and spin-flip opaque reflections: their kinematics and amplitudes. Section IV gives our main results. In Section V we discuss our approach as well as those of other authors. The paper is concluded by a Summary.
\section{\label{sec:level1} Spin splitting due to potential barrier\protect\\ \lowercase{}}

We first calculate the spin splitting of electron energies caused by the spin-orbit interaction between the incoming electron and the potential barrier. The interaction is of the general form given by Eq. (1) but the problem in our case is not trivial since the energy spectra of unperturbed and perturbed energies are continuous. We consider a finite potential barrier at $z = 0$. The starting point for the theory is the multiband {\kp}
formulation including the potential $V(\textbf{r})$ on the diagonal
\begin{equation}
\sum_l
\left[\left(-E+V+\varepsilon_{l0}+\frac{p^2}{2m_0}\right)\delta_{l'l} +\frac{1}{m_0}\textbf{p}_{l'l} \cdot \textbf{\textbf{p}} \right] f_l(\textbf{r}) = 0\;\;\;,
\end{equation}
where l'=1, 2, ..., $E$ is the electron energy, $\varepsilon_{l0}$ are the
band edge energies, $m_0$ is the free electron mass, $f_l(\textbf{r})$ are the envelope functions, and $\textbf{p}_{l'l}$ are the
interband matrix elements of momentum taken between the Luttinger-Kohn periodic amplitudes $u_l(\textbf{r})$. The potential
$V(\textbf{r})$ and the envelope functions $f_l(\textbf{r})$ are assumed to be slowly varying within the unit cell (which is not
always a realistic assumption). If the potential varies only in one dimension: $V(\textbf{r}) = V(z)$, one may separate the variables
by looking for solutions in the form $f_l(\textbf{r}) = exp(ik_x x+ik_y y) \chi_l(z)$. This gives $p_x\rightarrow \hbar k_x$,
$p_y\rightarrow \hbar k_y$, and the only nontrival variable is $z$. Within the three-level model (3LM) one deals with 8 bands arising from $\Gamma_6^c$,
$\Gamma_8^v$ (double degenerate) and $\Gamma_7^v$ levels. Thus the sum over $l$ in Eq. (2) runs from 1 to 8 and $l'$
=1...8, so that one deals with 8 coupled differential
equations for the functions $\chi_l(z)$. The parameters in set (2) are the band edge energies and the momentum matrix elements. Within 3LM there is one nonvanishing matrix element $P_0$ between $\Gamma^c_6$ and $\Gamma^v_8, \Gamma^v_7$ symmetry functions and one spin-orbit energy $\Delta$ splitting the $\Gamma^v_8$ and $\Gamma^v_7$ functions.
Using minor approximations one reduces set (2) by
substitution to two equations for the effective spin-up and spin-down conduction states, see Ref. 16. The resulting eigenvalue problem for the growth direction $z$ parallel to [001] crystal axis is
\ \\
\begin{equation}
 \left( \begin{array}{cc} \hat{A}- E&\hat{K}_{SIA} \\
\hat{K}^\dagger_{SIA}&\hat{A}- E \end{array} \right)
 \left( \begin{array}{c} \chi_1(z) \\ \chi_2(z) \end{array} \right) =0,
\end{equation}
\ \\
where
\ \\
\begin{equation}
\hat{A}=-\frac{\hbar^2}{2}\frac{\partial}{\partial z}
\frac{1}{m^*}\frac{\partial}{\partial
z}+\frac{\hbar^2k^2_\perp}{2m^*}+V(z)\;\;,
\end{equation}
\ \\
in which $k^2_{\perp} = k^2_x + k^2_y$ and $m^*(E, z)$ is the effective mass for the conduction electrons
\ \\
\begin{equation}
\frac{m_0}{m^*(E, z)}=1+C-\frac{E_P}{3}(\frac{2}{\tilde
\varepsilon_i}+ \frac{1}{\tilde f_i})\;\;,
\end{equation}
\ \\
where we explicitly indicate by the subscripts that $\tilde
\varepsilon_i$ and $\tilde f_i$ are different in various parts of the system.
Further
\ \\
\begin{equation}
\hat{K}_{SIA}=\frac{-ik_-}{\sqrt 2}\frac{\partial \eta(z)}{\partial
z}\;\;,
\end{equation}
where
\begin{equation}
\eta(z)=\frac{2P_0^2}{3}(\frac{1}{\tilde
\varepsilon_i}-\frac{1}{\tilde f_i})\;\;,
\end{equation}
\ \\
in which $k_- = (k_x - ik_y)/{\sqrt 2}$, while $\tilde \varepsilon_i(z) = \varepsilon_i(z)+V(z)-E$ and
$\tilde f_i(z) = \varepsilon_i(z)+\Delta_i(z)+V(z)-E$. Here $\varepsilon_i(z)$ are the energy gaps, $\Delta_i(z)$ are the
spin-orbit energies, $E_P = P^2_0 2m_0/\hbar^2$ and $C$
represents far-band contributions to the effective mass. The functions $\tilde
\varepsilon_i(z)$ and $\tilde f_i(z)$ depend on $z$ not only via
$V(z)$, but also due to the jumps of $\varepsilon_i$ and
$\Delta_i$ at the interfaces. We consider the case of steep barrier: $V(z) = 0$, $\varepsilon_i(z) = \varepsilon_A$, $\Delta_i(z) = \Delta_A$ and $\eta(z) = \eta_A$ for $z\le 0$, while $V(z) = V_b$, $\varepsilon_i(z) = \varepsilon_B$, $\Delta_i(z) = \Delta_B$ and $\eta(z) = \eta_B$ for $z > 0$, see Fig. 1. We denote $m^*(E, z) = m^*_A$ on the left of the barrier and $m^*(E, z) = m^*_B$ on the right. We consider an electron coming to the barrier from an oblique direction. Without loss of generality one can choose the coordinate system in such a way that $k_y = 0$, while $k_x \ne 0$ and $k_z \ne 0$.

It is seen that for $K_{SIA} = 0$ the two spin states have the same energy. We will solve set (1) using the well known approximate procedure for the perturbation of two degenerate states by calculating the matrix element of the nondiagonal term $K^{\uparrow\downarrow}_{SIA}$ between the solutions of diagonal terms. Explicitly
\ \\
\begin{equation}
E_{1, 2}=\hat{A}^{\uparrow\uparrow} \pm|\hat{K}_{SIA}^{\uparrow\downarrow}|\;\;.
\end{equation}
\ \\
The matrix element is taken between two orthogonal spin states $\chi_{\uparrow}(z) = (1\;,\;0)^T \chi(z)$ and $\chi_{\downarrow}(z) = (0\;,\;1)^T \chi(z)$, where $\chi(z)$ is solution of the Schrodinger equation $\hat{A}\chi(z) = E\chi(z)$, see Eq. (3). The unperturbed and perturbed electron spectrum is continuous and the wave functions are plane waves. This means that the wave functions along the $z$ direction must be normalized to the delta function $\delta(k_z - k'_z)$. In consequence, the matrix elements $A^{\uparrow\uparrow}$ and $K^{\uparrow\downarrow}_{SIA}$ have the dimension of [energy $\times$ length]. In order to obtain the correct dimension of [energy] one has to integrate Eq. (8) over $k'_z$. There is, for each spin, an incoming and a reflected plane wave for $z \le 0$ and a decaying wave for $z > 0$. The incoming wave vector is $k_z > 0$, the reflected one is $-k_z$ and the decaying wave is described by the imaginary wave vector $q_z$. The value of $k_x$ does not change since there is no force acting along the $x$ direction. Explicitly, the unperturbed spin-up and spin-down wave functions have the standard form
\ \\
\begin{equation}
\chi_{\uparrow} = \frac{\sqrt{N}}{\sqrt{2\pi L_x}} e^{ik_x x }\left(\begin{array}{c}1\\0\end{array}\right) \chi(z)
\;\;,
\end{equation}
\\
\begin{equation}
\chi_{\downarrow} = \frac{\sqrt{N}}{\sqrt{2\pi L_x}} e^{ik_x x }\left(\begin{array}{c}0\\1\end{array}\right) \chi(z)
\;\;,
\end{equation}
\ \\
where
\ \\
\begin{equation}
\chi(z) = (e^{ik_z z} + R e^{-ik_z z})_{| z\le 0} + (T e^{iq_z z})_{| z > 0}
\;\;.
\end{equation}
\ \\
Along the $x$ direction the functions are normalized to the length $L_x$, while along the $z$ direction we aim at the normalization to the delta
function $\delta(k_z - k'_z)$ and keep the normalization coefficient $\sqrt N$, where $N$ is to be determined.

We first consider the boundary conditions
at the interface. There is
\ \\
\begin{equation}
\chi|_{+_{0}}=\chi|_{-_{0}}\;\;\;,
\end{equation}
\ \\
and
\ \\
\begin{equation}
\left(\frac{1}{m^*_{B}}\frac{\partial \chi}{\partial
z}\right)_{+_{0}} = \left(\frac{1}{m^*_A}\frac{\partial
\chi}{\partial z}\right)_{-_{0}}\;\;\;.
\end{equation}
\ \\
This gives for both spin functions
\ \\
\begin{equation}
T = 1+R\;\;\;,
\end{equation}
\ \\
\begin{equation}
\frac{ik_z(1-R)}{m^*_A} = \frac{iTq_z}{m^*_B}\;\;\;,
\end{equation}
\ \\
and finally
\ \\
\begin{equation}
R=\frac{(k_z-q_z M)}{(k_z+q_z M)}\;\;\;.
\end{equation}
\ \\
where $ M = m^*_A/m^*_B$.
\ \\

We assume that initially the electrons are characterized by the energy $E$ and the wave vector $k_x$. Since in the region
 A there is $V(z) = 0$, the energy of incoming electrons is $E = \hbar^2(k^2_x + k^2_z)/2m^*_A$, so we have
\ \\
\begin{equation}
k_z^2=\left(E\frac{2m^*_A}{\hbar^2}-k^2_x\right)\;\;\;,
\end{equation}
\ \\
while in the barrier region B there is
\ \\
\begin{equation}
q_z^2=\left[(E-V_b)\frac{2m^*_B}{\hbar^2}- k_x^2\right]\;\;\;.
\end{equation}

Now we calculate the normalization of $\chi(z)$.
\ \\
$$
\frac{N}{2\pi}\int_0^{\infty}dk'_z\int_{-\infty}^0dz (e^{ik_z z}+Re^{-ik_z z})(e^{-ik'_z z} + R'^*e^{ik'_z z}) +
$$
\ \\
\begin{equation}
+ \frac{N}{2\pi}\int_0^{\infty}dq'_z\int_0^{\infty}dz Te^{iq_z} T'^*e^{-iq'_z z} = 1\;\;.
\end{equation}
\ \\
The final formula for $N$ depends on the relation between the electron energy $E$ and $V_b + \hbar^2k_x^2/2m^*_B$.

a) The first case is: $E < V_b + \hbar^2k_x^2/2m^*_B$, so that $q_z = i|q_z|$ according to Eq. (18). The function in the barrier $Te^{-|q_z|z}$ decays fast and we can neglect the term proportional to $T$. Consequently, $\chi(z)$ is nonzero for $-\infty < z \le 0$. The normalization condition is
\ \\
$$
\frac{N}{2}\int_0^{\infty} dk'_z\left( (1 + R R^{'*})\delta(k_z - k'_z)+\right.
$$
\begin{equation}
\left. - (R + R^{'*})\delta(k_z + k'_z)\right) = 1
\end{equation}
\ \\
The second term gives no contribution to the integral since both $k_z$, $k'_z > 0$. The first term imposes $k'_z = k_z$, then from Eq. (16) there is $1 + R R'^* = 2$, so that finally $N = 1$.
It should be noted that the integration over $k'_z$ is related to the correct dimension of energy in Eq. (3), as mentioned above.

b) The second case is, $E > V_b + \hbar^2k_x^2/2m^*_B$, so that $q_z, R, T$ are real. The function
$\chi(z)$ is nonzero for $-\infty < z < \infty$. The normalization condition is
\ \\
$$
\frac{N}{2}\left[\int_0^{\infty} dk'_z (1 + R R')\delta(k_z - k'_z) + \right.
$$
\begin{equation}
\left. +\int_0^{\infty} dq'_z T T'\delta(q_z - q'_z)\right] = 1\;\;.
\end{equation}
\ \\
The first term gives $1+R^2$ and the second one gives $T^2 = (1+R)^2$. Finally there is $N = (k_z+Mq_z)^2/(3k_z^2+M^2q_z^2)$.
\begin{figure}
\includegraphics[scale=0.45,angle=0, bb = 700 20 202 540]{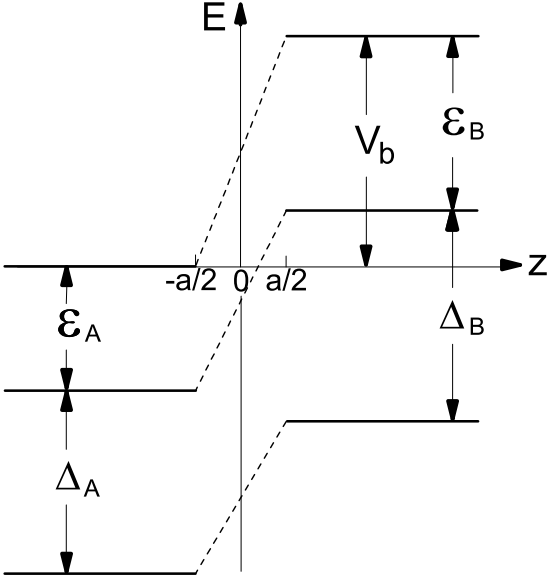}
\caption{\label{fig:epsart}{One-dimensional potential barrier used for the calculation of spin-orbit energy (schematically). Energy gaps and spin-orbit energies in regions A (InSb) and B (InAlSb) are indicated.}} \label{fig1th}
\end{figure}
\ \\

To calculate the spin splitting determined by the average value of $|\hat{K}^{\uparrow \downarrow}_{SIA}|$ = $\Delta$, we take into account the
valence offsets contained in the $\tilde \varepsilon_i(z)$ and $\tilde f_i(z)$ functions. We will assume here that the potential barrier is a steep but linear function of $z$, extending from  $-a/2$ to $a/2$, see Fig. 1. After some manipulation the expression for $\Delta$ is brought to the form
$$
\Delta = \int^{\infty}_0 dk'_z|\int^{a/2}_{-a/2}dz \chi^*_{\downarrow}(z,k'_z)\hat{K}_{SIA}\chi_{\uparrow}(z,k_z)|  =
$$
\ \\
$$
= \frac{Nk_x}{2}\frac{(\eta_B-\eta_A)}{a}
\int^{\infty}_0dk'_z [1+R^*(k'_z)][1+R(k_z)]\cdot
$$
$$
\cdot\frac{2}{2\pi}\frac{\sin[(k_z-k'_z)a/2]}
{(k_z-k'_z)} =
$$
\ \\
$$
=\frac{Nk_x}{2}\frac{(\eta_B-\eta_A)}{a}\int_0^{\infty}[1+R^*(K'_z)][1+R(K_z)]\cdot
$$
\begin{equation}
\cdot\frac{\sin[(K_z-K'_z)D]}{\pi(K_z-K'_z)}dK'_z
\end{equation}
\ \\
By changing the variable $k_z$ into $K_z = k_z a/2D$, $k'_z$ into $K'_z = k'_z a/2D$, $|q_z|M$ into $|Q_z| = |q_z| M a/2D$ and $|q'_z|$ into $|Q'_z| = |q'_z| M a/2D$, where $D$ is a large number and $R(K_z) = ((K_z-Q_z)(K_z+Q_z)$, the last factor in the integrand of Eq. (22) can be replaced to a very good approximation by $\delta(K_z - K'_z)$. Then the integration over $dK'_z$ in Eq. (22) can be  carried out and the final result for the case $E < V_b + \hbar^2k_x^2/2m^*_B$ is
\begin{equation}
\Delta_{E <} = \frac{k_x(\eta_B-\eta_A)}{2a}
\frac{4k_z^2}{(k_z^2 + |q_z|^2 M^2)}\;\;,
\end{equation}
\ \\
while for the case $E > V_b + \hbar^2k_x^2/2m^*_B$ there is
\begin{equation}
\Delta_{E >} = \frac{k_x(\eta_B-\eta_A)}{2a}
\frac{4k_z^2}{(3k_z^2 + q_z^2 M^2)}\;\;.
\end{equation}
\ \\
The spin splitting is $2\Delta$, see Eq. (8) and Fig. 2.

It can be
seen that the spin splitting of the conduction band due to the SIA mechanism is proportional to the spin-orbit energies in the
valence bands contained in $\eta_B - \eta_A$ term. Also, $\Delta$ depends on $k_z$ and $q_z$, which characterize the incoming electrons. In the published literature one deals with the Bychkov-Rashba spin splitting in the quantum wells, where $\Delta$ is related to the electron wave function in the well. The unusual feature of our situation is that we do not deal with the well, so the wave function is replaced by the incoming and returning waves. According to the two-state perturbation procedure, as formulated in Eqs. (3) and (8), the resulting wave functions for the effective spin-up and spin-down have the form $(1, 1)^T$ and $(1, -1)^T$, respectively.
\section{\label{sec:level1} Spin-conserving and spin-flip reflection\protect\\ \lowercase{}}

The off-diagonal terms $\hat{K}_{SIA}$ mix the spins in the wave functions due to the SOI in the complete solutions of Eq. (3). The resulting spin-mixed functions correspond to the \emph{effective} spin-up and spin-down states, which we label by subscripts $1$ and $2$, respectively.
\begin{widetext}
\begin{equation}
\Psi_1=\frac{e^{ik_x x}}{\sqrt{4\pi L_x}}\left\{\left[e^{ik'_z z}\left(\begin{array}{c}1\\1\end{array}\right)+Re^{-ik'_z z}\left(\begin{array}{c}1\\1\end{array}\right)+R'e^{-ik''_z z}\left(\begin{array}{c}1\\-1\end{array}\right)\right]_{|z\le 0}+
\left[Te^{iq'_z z}\left(\begin{array}{c}1\\1\end{array}\right)+T'e^{iq''_z z}\left(\begin{array}{c}1\\-1\end{array}\right)\right]_{|z>0}\right\}
\;,
\end{equation}
\ \\
\begin{equation}
\Psi_2=\frac{e^{ik_x x}}{\sqrt{4\pi L_x}}\left\{\left[e^{ik''_z z}\left(\begin{array}{c}1\\-1\end{array}\right)+
Pe^{-ik''_z z}\left(\begin{array}{c}1\\-1\end{array}\right)+P' e^{-ik'_z z}\left(\begin{array}{c}1\\1\end{array}\right)\right]_{|z\le0}+
\left[Fe^{iq''_z z}\left(\begin{array}{c}1\\-1\end{array}\right)+F'e^{iq'_z z}\left(\begin{array}{c}1\\1\end{array}\right)\right]_{|z>0}\right\}
\;,
\end{equation}
\end{widetext}
\ \\
Function $\Psi_1$ describes a spin-up electron with the incoming wave vector $k'_z > 0$ which, if reflected without the change of spin, is returning with the wave vector $-k'_z$ and has the penetrating component characterized by $q'_z$. If the electron is reflected with the change of spin direction, the returning wave is characterized by $-k''_z$ and the penetrating one by $q''_z$. Things are analogous for the electron with the initial effective spin down.
Suppose the incoming electron is in the state $\Psi_1$, exactly speaking in the state $(1, 1)^T$, so that its energy is
\ \\
\begin{equation}
E_1 = \frac{\hbar^2 {k'}^2}{2m^*_A} + \Delta
\;\;,
\end{equation}
\ \\
where ${k'}^2 = k_x^2 + {k'}_z^2$. After an elastic reflection from the barrier there are two possibilities. If the reflected electron is still in the $(1, 1)^T$ state (spin-conserving process) its energy is again given by Eq. (27), the resulting $k_x$ remains the same while $k'_z$ changes sign. This means that, in a spin-conserving reflection, the outgoing and incoming electron directions form the same angles with the normal to the barrier. If, on the other hand, the reflected electron is in the $(1, -1)^T$ state (spin-flip process), its energy is
\ \\
\begin{equation}
E_2 = \frac{\hbar^2 {k''}^2}{2m^*_A} - \Delta
\;\;.
\end{equation}
\ \\
where ${k''}^2 = k_x^2 + {k''}_z^2$. In an elastic reflection the total electron energy must remain the same, so we have from Eqs. (27) and (28)
\begin{equation}
\frac{\hbar^2}{2m^*_A}({k''}^2 - {k'}^2) = 2\Delta\;\;.
\end{equation}
\begin{figure}
\includegraphics[scale=0.45,angle=0, bb = 700 20 202 540]{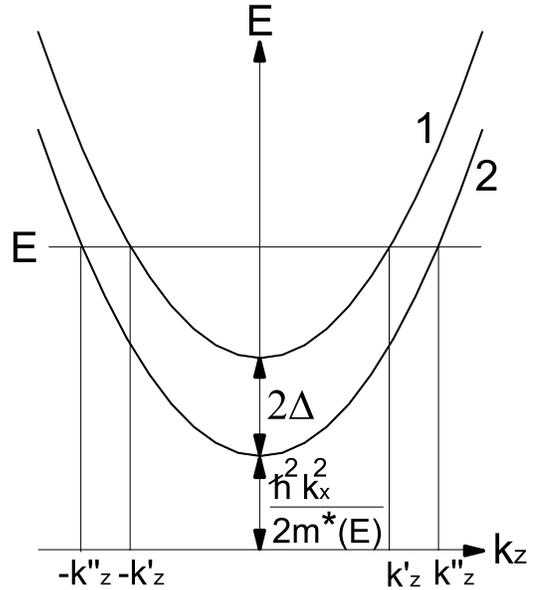}
\caption{\label{fig:epsart}{Electron energy versus wave vector $k_z$ for spin-up and spin-down states (schematically). The spin-orbit energy is $\Delta$. For a spin-conserving elastic reflection of spin-up electron the outgoing $k'_z$ is equal to $-k'_z$, for a spin-flip reflection the outgoing wave vector is $-k''_z$ such, that $|k''_z| > |k'_z|$. Situation for an initial spin-down electron with the initial wave vector $k''_z$ is also shown.}} \label{fig2th}
\end{figure}
Since $k_x$ does not change, there is
\begin{equation}
{k'}_z^2 = (E - \Delta)\frac{2m^*_A}{\hbar^2} - k_x^2\;\;,
\end{equation}
\ \\
and
\begin{equation}
{k''}_z^2 = (E + \Delta)\frac{2m^*_A}{\hbar^2} - k_x^2\;\;.
\end{equation}
\ \\
This means that, because of the spin splitting related to the SOI, in a spin-flip reflection the outgoing and incoming directions do \emph{not} form the same angles with the normal to the barrier. The above reasoning is illustrated in Fig. 2. Clearly, there are limits to the initial and final directions: they should not be parallel to the barrier. The incoming angle with the normal to the barrier is
$\alpha'$ = ctg$^{-1}(k'_z/k_x)$ and the outgoing angle after spin-flip process is: $\alpha''$ = ctg$^{-1}(k''_z/k_x)$ and the difference of angles between spin-conserving and spin-flip reflected electron beams is $\alpha' - \alpha''$.

Now we turn to the amplitudes of functions (25) and (26). Let us again consider an incoming spin-up electron. There exist, in addition to the waves characterized by $k'_z$ and $k''_z$ considered above, waves penetrating the barrier. For the spin-up penetrating wave the wave vector $q'_z$ is
\begin{equation}
q'^2_z=(E - \Delta - V_b)\frac{2m_B}{\hbar^2}-k^2_x\;\;,
\end{equation}
\ \\
while for the spin-down penetrating wave the wave vector $q''_z$ is
\begin{equation}
q''^2_z=(E + \Delta - V_b)\frac{2m_B}{\hbar^2}-k^2_x\;\;\;.
\end{equation}
\ \\
The spin-conserving and spin-flip waves penetrating the barrier form the angles $\beta'$ = ctg$^{-1}(|q'_z|/k_x)$ and $\beta''$ = ctg$^{-1}(|q''_z|/k_x)$ with the normal to the step potential, respectively, see Fig. 3.
\begin{figure}
\includegraphics[scale=0.45,angle=0, bb = 600 20 202 540]{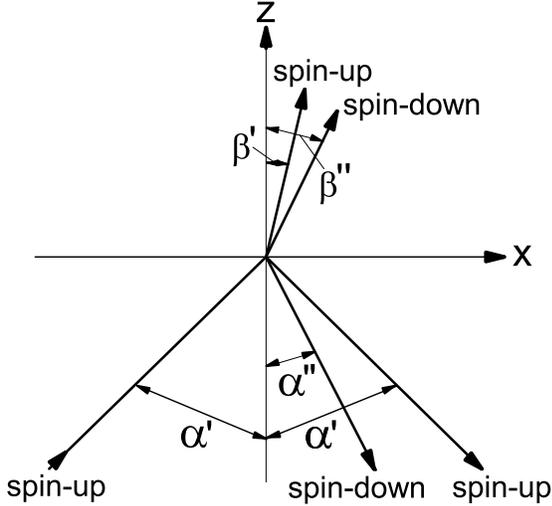}
\caption{\label{fig:epsart}{Geometry of spin-polarized electrons reflected from a
semiconductor potential barrier (schematically). In a spin-conserving reflection of a spin-up electron the
outgoing angle is equal to the incoming angle $\alpha'$, in a spin-flip
reflection the outgoing angle $\alpha''$ is smaller: $\alpha'' < \alpha'$. The waves penetrating the barrier at angles $\beta'$ and $\beta''$ are also shown. The effect is due to the spin-orbit interaction manifested in
oblique motion, see Fig. 1.
}} \label{fig3th}
\end{figure}
In the following we consider explicitly an incoming spin-up electron characterized by $\Psi_1$.
The amplitudes of incoming, reflected and penetrating parts of the function can be determined from the boundary conditions for the wave functions and their derivatives at $z = 0$. For the upper-spin components in Eq. (27) the continuity condition is
\ \\
\begin{equation}
 {\Psi_1^{up}}_{| z = 0^-} = {\Psi_1^{up}}_{| z = 0^+}\;\;,
\end{equation}
\ \\
while for the lower spin components there is
\begin{equation}
{\Psi_1^{low}}_{| z = 0^-} = {\Psi_1^{low}}_{| z = 0^+}\;\;\;.
\end{equation}
\ \\
The remaining boundary conditions come from the integration of Eqs. (3) across the interface at $z = 0$. We have
\ \\
$$
\frac{\hbar^2}{2m^*_A}\frac{\partial {\Psi_1^{up}}}{\partial z}_{| z = 0^-} = \frac{\hbar^2}{2m^*_B}\frac{\partial {\Psi_1^{up}}}{\partial z}_{| z = 0^+} +
$$
\ \\
\begin{equation}
i\frac{k_x(\eta_B-\eta_A)}{2}{\Psi_1^{low}}_{| z = 0}\;\;\;,
\end{equation}
\ \\
$$
\frac{\hbar^2}{2m^*_A}\frac{\partial {\Psi_1^{low}}}{\partial z}_{| z = 0^-} = \frac{\hbar^2}{2m^*_B}\frac{\partial {\Psi_1^{low}}}{\partial z}_{| z = 0^+} - $$
\ \\
\begin{equation}
i\frac{k_x(\eta_B-\eta_A)}{2}{\Psi_1^{up}}_{| z = 0}\;\;\;,
\end{equation}
\ \\
which gives
\begin{equation}
 \frac{\partial {\Psi_1^{up}}}{\partial z}_{| z = 0^-} = \frac{m^*_A}{m^*_B}\frac{\partial {\Psi_1^{up}}}{\partial z}_{| z = 0^+} + iS{\Psi_1^{low}}_{| z = 0}\;\;\;,
\end{equation}
\ \\
\begin{equation}
 \frac{\partial {\Psi_1^{low}}}{\partial z}_{| z = 0^-} = \frac{m^*_A}{m^*_B}\frac{\partial {\Psi_1^{low}}}{\partial z}_{| z = 0^+} - iS
{\Psi_1^{up}}_{| z = 0}\;\;\;,
\end{equation}
\ \\
where $S = k_x(\eta_B-\eta_A)m^*_A/\hbar^2$. The terms with $\partial \Psi_1^{up/low}/\partial z$ come from the integration of the diagonal terms in Eq. (3), while the terms with $S$ result from the integration of the nondiagonal terms related to the SOI. We set, as before, $k_y = 0$.
From Eqs. (34) and (35) one gets have: $1 + R + R' = T + T'$ and $1 - R - R'= T - T'$, which gives $R' = T'$ and $R = T - 1$.
From Eqs. (38) and (39) we have
\ \\
$$
ik'_z(1-R)-ik''_zR'=
$$
\begin{equation}
iMq'_zT+iMq''_zT'+iS(T-T')\;\;\;,
\end{equation}
\ \\
$$
 ik'_z(1-R)+ik''_zR'=
 $$
 \begin{equation}
 iMq'_zT-iMq''_zT'-iS(T+T')
\;\;\;,
\end{equation}
\ \\
where $M = m^*_A/m^*_B$. After some manipulations we finally obtain from Eqs. (40) and (41)
\ \\
\begin{equation}
R = \frac{[(k'_z-Mq'_z)(k''_z + Mq''_z)-S^2]}{[(k'_z+Mq'_z)(k''_z+Mq''_z)+S^2]}
\end{equation}
\ \\
\begin{equation}
R' = \frac{-2S k'_z}{[(k'_z+Mq'_z)(k''_z+Mq''_z)+S^2]}
\end{equation}
\ \\
\begin{equation}
T  = \frac{2k'_z(k''_z+Mq''_z)}{[(k'_z+Mq'_z)(k''_z+Mq''_z)+S^2]}
\;\;\;,
\end{equation}
\ \\
\begin{equation}
T' = R'\;\;\;.
\end{equation}
\ \\
It is seen that the amplitudes of spin-flip reflected and penetrating waves, proportional to $R'$ and $T'$ terms in Eq. (25), do not vanish because of the $S$ terms in Eqs. (43) and (45). The latter are due to the spin-orbit energies in the valence bands and, technically, come about from the spin-dependent boundary conditions for $\Psi_1$ function in $z = 0$.

A similar analysis can be carried out for the incoming spin-down electron with the use of Eq. (26). For such electrons one has $P = R$, $P' = -R'$, $F = T$ and $F' = -T'$. At the same time one should replace in the formulas $k'_z$ by $k''_z$, $q'_z$ by $q''_z$, $k''_z$ by $k'_z$, and $q''_z$ by $q'_z$. If we use Fig. 3 then, qualitatively speaking, for the initial spin-up electron, the spin-flip process results in a reflection angle smaller than the incoming one, while for the initial spin-down electron the spin-flip process results in a larger reflection angle than the incoming one.

Since, as follows from Eqs. (23 - 24), the spin splitting 2$\Delta$ is characterized by $k_x$, $k_z$ and $q_z$ and, in turn, the kinematics in determined by $\Delta$, one should analyze how the final results are related to the initial conditions. Suppose one begins with the spin-up electron having the energy $E$ and coming to the barrier at the angle $\alpha'$. Then one gets $k_x = k'_z (tg{\alpha'})$. One can now use Eq. (17) to obtain $k_z^2 = \left(E\frac{2m^*_A}{\hbar^2}-k'^2_z tg^2{\alpha'}\right)$ and Eq. (18) to get $q_z^2 = \left[(E-V_b)\frac{2m^*_B}{\hbar^2}- k'^2_z tg^2{\alpha'}\right]$. Putting the above expressions for $k_x$, $k_z$, $q_z$ to Eq. (23) for $\Delta_{E < V_b}$ or to Eq. (24) in order to calculate $\Delta_{E > V_b}$, and using Eq. (30) we obtain an equation for one unknown $k'_z$. Having determined $k'_z$ we calculate $k_x$, $k_z$ $q_z$, then $\Delta$ and finally $k''_z$, $q'_z$ and $q''_z$. This way one can calculate the angle $\alpha''$ and then the wave amplitudes given by the above formulas.
\section{\label{sec:level1} Results\protect\\ \lowercase{}}

When performing experiments one can imagine two different situations. In the first, a source produces spin-polarized electrons which come to the barrier from an opaque direction and are reflected in the spin-conserving or spin-flip processes. This case is illustrated in Fig. 3. In the second situation, realized in the experiments of Chen et al [10, 12, 14], electrons come to the barrier from a defined direction but they are not spin polarized. In this case one can consider the spin to be a combination of spin-up and spin-down components. For the spin-flip processes, the reflected spin-down component has the direction further from normal (see Fig. 3), the spin-up component has the direction closer to the normal, and the spin-conserving processes give the reflection having the same angle as the incoming beam. Thus the reflected beams on both sides contain spin polarized electrons while the middle beam contains unpolarized electrons.
\begin{table}[h]
\caption{Material parameters used for InSb/In$_{0.91}$ Al$_{0.09}$Sb barrier. The lattice constant $a$ is taken as an average value between InSb and In$_{0.91}$Al$_{0.09}$Sb.}
\begin{tabular}{c|c|c|c|c}

\hline &&InSb&In$_{0.91}$Al$_{0.09}$Sb&\\
\hline
\hline
&$E_g$(eV)& -0.2466 & -0.426&\\
&$\Delta_0$(eV)& -0.8419 & -0.777 &\\
&$m^*_0/m_0$& 0.0143 & 0.0234 &\\
&C& -1.496 & -1.3685 &\\
&$E_p$(eV)& 23.4 & 23.4 &\\
&$a$($10^8$cm)& 6.4&6.4&\\
&$V_b$(eV)& - & 0.111 &\\\hline
\end{tabular}
\end{table}
\ \\
\begin{table}[h]
\caption{Heterojunction parameters used for InSb/In$_{0.91}$ Al$_{0.09}$Sb barrier. The effective masses $m^*_A/m_0$ and $m^*_B/m_0$ are calculated for given electron energies $E$.}
\begin{tabular}{c|c|c|c|c}
\hline &&InSb&In$_{0.91}$Al$_{0.09}$Sb&\\
\hline
\hline
&E(eV)& 0.08 & 0.08 &\\
&$m^*_A/m_0$&0.01854 & - &\\
&$m^*_B/m_0$& - & 0.02184 &\\ \hline
&E(eV)& 0.13 & 0.13 &\\
&$m^*_A/m_0$& 0.02113 & - &\\
&$m^*_B/m_0$& - & 0.02435 &\\\hline
\end{tabular}
\end{table}

Now we estimate numerical values of the reflected directions and amplitudes by considering specific barrier: InSb/In$_{0.91}$Al$_{0.09}$Sb. In Table I we quote the band parameters of the involved narrow-gap semiconductor materials plus the offset $V_b$ of the heterojunction. All the other parameters can be inferred from the given values. In Table II we quote the effective masses calculated for the electron energies $E$ in regions A and B. In Tables III and IV we quote  characteristics of reflected and penetrating waves, as calculated for different initial sets of parameters. The initial sets are the electron energy $E$ for $E < V_b + \hbar^2k_x^2/2m^*_B$ (Table III) and $E > V_b + \hbar^2k_x^2/2m^*_B$ (Table IV)) and the wave vector $k_x$ (two values) determining the incoming angle $\alpha'$ (for spin-up electrons) and $\alpha''$ (for spin-down electrons). In entries for $k_x = 0.35 \cdot 10^6$ cm$^{-1}$ we deal with the "standard" case in which $k'_z$ and $k''_z$ have real values, while $q'_z$ and $q''_z$ have imaginary values for $E < V_b + \hbar^2k_x^2/2m^*_B$ and real values for $E > V_b + \hbar^2k_x^2/2m^*_B$. An interesting and unexpected situation is described in the entries for $k_x = 1\cdot10^6$ cm$^{-1}$. It follows from Tables I and III that, for the incoming energy $E$ = 80 meV of the spin-down electron and $V_b$ = 111 meV, there is $E < V_b + \hbar^2k_x^2/2m^*_B$, i.e. the kinetic electron energy is smaller than the barrier height. According to the intuitive expectations, it should be a "normal" case in which the wave penetrating the barrier quickly decays with the distance $z$. However, the calculated spin-orbit energy for this case is $\Delta$ = 58 meV, so that $E + \Delta$ = 138 meV is higher than $V_b$. As a consequence, it follows from Eq. (33) that the resulting $q''_z$ = 0.736 $\cdot 10^6$ cm$^{-1}$ is \emph{real}. This means that the penetrating spin-down wave $exp(iq''_z z)$\emph{ propagates in the barrier without attenuation}!
\begin{table}[h]
\caption{Case $E < V_b + \hbar^2k_x^2/2m^*_B$. Spin-orbit energies and reflection characteristics for spin-up electron ($\alpha'$) and spin-down electron ($\alpha''$) calculated for $E$ = 0.08 eV and $V_b$ = 0.111 eV, and given values of $k_x$.}
\begin{tabular}{c|c|c|c|c}
\hline
&$k_x$($10^{-6}$/cm)& 0.35&1.0&\\
&$\Delta$(eV)& 0.025 &0.058&\\
\hline
\hline
&$\alpha'$& 12.37$^o$ &74.63$^o$ &\\
&$q'_z$($10^{-6}$/cm)&i1.828 &i2.47& \\
&$R$&0.028-i0.999&=0.97-i0.24& \\
&$R'$&-0.008+i0.014&i0.006& \\
&$RC_1$& 1&0.9994& \\
&$TC_1$& 0&0.0006& \\
\hline
&$\alpha''$& 8.90$^o$ &22.71$^o$ &\\
&$q''_z(10^{-6}$/cm)&i0.676&0.736&\\
&$P$& 0.89-i0.45 &0.585& \\
&$P'$& 0.013-i0.02 &0.006-i0.05& \\
&$RC_2$&1&0.343& \\
&$TC_2$&0&0.657& \\
\hline
\end{tabular}
\end{table}
\ \\
\begin{table}[h]
\caption{Case $E > V_b + \hbar^2k_x^2/2m^*_B$. The same as in Table III but for $E$ = 0.13 eV and $V_b$ = 0.111 eV.}
\begin{tabular}{c|c|c|c|c}
\hline
&$k_x(10^{-6}$/cm)& 0.35&1.0&\\
&$\Delta$(eV)& 0.0085 & 0.025 &\\
\hline
\hline
&$\alpha'$& 7.75$^o$&24.48$^o$&\\
&$q'_z(10^{-6}$/cm)&0.74&i1.176& \\
&$R$&-0.600&0.645-i0.764& \\
&$R'$& -0.0098&-0.03+i0.01& \\
&$RC_1$&0.361&0.9997& \\
&$TC_1$&0.639&0.0003& \\
\hline
&$\alpha''$&7.25$^o$&19.94$^o$ &\\
&$q''_z(10^{-6}$/cm)&1.279&1.346&\\
&$P$& 0.425&0.405& \\
&$P'$&0.0104&0.035-i0.016& \\
&$RC_2$&0.18&0.164& \\
&$TC_2$&0.82&0.836& \\
\hline
\end{tabular}
\end{table}

An "inverse" situation is created by the incoming spin-up electron with the energy $E$ = 130 meV, see Table IV. For the barrier $V_b$ = 111 meV there is $E > V_b + \hbar^2k_x^2/2m^*_B$, so that the incoming electron moves in principle above the barrier. However, for this case there is $\Delta$ = 25 meV. So that $E-\Delta < V_b + \hbar^2k_x^2/2m^*_B$ and, according to Eq. (32), the resulting $q'_z$ is imaginary. This means that the continuing spin-up component \emph{quickly decays above the barrier} and there is an almost total reflection of the spin-up wave.

It is instructive to calculate not only the amplitudes, but also reflection and transmission coefficients of various waves. A transmission coefficient TC is defined as TC=$|J_{tr}/J_{in}|$ and the reflection coefficient as RC=$|J_{re}/J_{in}|$, where $J_{in}$ is the incoming current.
According to Eqs. (25 - 26), we have for the incoming spin-up electron
\begin{equation}
TC_1 = (|T|^2q'_z + |T'|^2q''_z)(\frac{ m^*_A}{k'_z m^*_B})\;\;,
\end{equation}
\ \\
\begin{equation}
RC_1 = |R|^2+|R'|^2\frac{k''_z}{k'_z}\;\;,
\end{equation}
\ \\
while for the incoming spin-down electron
\begin{equation}
TC_2 = (|F|^2q''_z + |F'|^2q'_z)(\frac{ m^*_A}{k''_z m^*_B})\;\;,
\end{equation}
\ \\
\begin{equation}
RC_2 = |P|^2+|P'|^2\frac{k'_z}{k''_z}\;\;,
\end{equation}
where the second terms in the above equalities correspond to spin-flip amplitudes.
For imaginary $q'_z$ or $q''_z$ we have the decaying waves in the barrier, i. e. $J_{tr}$ = 0 and TC = 0.
In the last entries of Tables III and IV we give the values of RC and TC, as calculated from the above formulas. It can be seen that RC + TC = 1, as it should be. One can also see that in the interesting two cases, i. e. the non-attenuated transmitted wave for $E < V_b + \hbar^2k_x^2/2m^*_B$ (Table III) has large transmission coefficient TC$_2$ = 0.657, while for electron with $E > V_b + \hbar^2k_x^2/2m^*_B$ (Table IV) the transmitted wave is attenuated in the barrier and has the transmission coefficient TC$_1$ = 0.0003. Small values of TC$_1$ = 0.0006 in Table III and TC$_1$ = 0.0003 in Table IV merit special comments. Concerning Table III, there is a small component of the spin-flip wave which is transmitted without attenuation (TC$_1$ = 0.0006), while the spin-conserved wave is completely reflected (RC$_1$ = 0.9994), so that the sum TC$_1$ + RC$_1$ = 1, as it should be. Similar reasoning applies to TC$_1$ = 0.0003 and RC$_1$ = 0.9997 in Table IV.

It is clear that the considered system can serve as a source or a filter of spin-polarized electron beams since it spatially separates electrons with specifically oriented effective spins. One can also say that this arrangement realizes the Stern-Gerlach experiment for mobile electrons, which is not possible for free electrons in a vacuum (see discussion in Ref. 15) and very difficult for electrons in semiconductors (Wrobel et al. [16]).
\section{\label{sec:level1} Discussion\protect\\ \lowercase{}}

We first examine the approximations used in our approach. Our description of the potential barrier is somewhat ambiguous since we sometimes use a vertical barrier and, when calculating the spin-orbit energy for unbound electrons, a barrier of finite slope. As we mentioned in Section II, the latter assumption was necessary in order to avoid an infinite electric field. However, this ambiguity is not troublesome since one needs a \emph{steep} barrier to obtain sizable spin splittings. On the other hand, a narrow barrier (on the order of interatomic distance) is in a certain contradiction with the validity of the {\kp} theory. The latter requires that the potential changes be slow compared to the lattice period. Fortunately, it is known that the {\kp} theory, usually gives better results than it is supposed to. Also, when calculating SOE we assume in Eq. (22) that the reflection amplitudes $R$ do not depend on $z$. All in all, it is possible that our calculations of SOE for unbound electrons are not very precise. This, however, should not influence qualitatively our results. To the best of our knowledge, a calculation of the Bychkov-Rashba spin splitting for unbound electrons has not been attempted before.

As it was mentioned in the Introduction, the spin-flip ballistic reflection of electrons from a potential barrier was experimentally realized by Chen and coworkers in InSb/InAlSb and InAs/AlGaSb heterostructures [10, 12, 14]. In their experiments the authors used a "triangular geometry": two sides of a triangle with slits were employed to inject and detect electron beams, the third side was utilized as a barrier. An external magnetic field transverse to the 2D plane of the heterostructure was used to direct the reflected beams to the detecting slit. A theoretical description of these experiments was presented mostly in Ref. 11. The authors assumed that the spin splitting of electron energies due to the SOI had two origins: the structural asymmetry of the quantum well and the barrier. However, it was then incorrectly concluded that the Bychkov-Rashba spin splitting of electron energies due to the asymmetric well was proportional to the electric field in the conduction band and that the spin splittings caused by the two origins were governed by the same material parameter. A calculation of spin splitting due to the SO interaction of unbound electrons with the barrier was not attempted. Teodorescu and Winkler [17] in their analysis of spin-dependent reflections from an impenetrable barrier used as an origin of the Bychkov-Rashba spin splitting only the inversion asymmetry of the well. Also in the work on the spin separation in 2D cyclotron motion (Ref. 18) and transverse electron focusing in 2D systems with the SOI (Ref. 19), the Bychkov-Rashba splitting for unbound electrons was taken into account only phenomenologically.

It should be emphasized that, as follows from our Eq. (23) and Ref. 2, the Bychkov-Rashba splitting for unbound and bound electrons bear some similarities but are also marked by important differences. In both cases it is necessary to have the SOI and structure inversion asymmetry. In both cases it is necessary to have a transverse motion and the resulting spin splitting is proportional to the transverse momentum. On the other hand, for the unbound electrons the spin splitting depends on the longitudinal momentum $\hbar k_z$, while for the bound electrons this momentum is "frozen" in the electron wave function of the quantum well and it does not explicitly appear. Also, as follows from Eqs. (23) and (24), for the unbound electrons even the functional forms of $\Delta$ are different for $E < V_b + \hbar^2k_x^2/2m^*_B$ and $E > V_b + \hbar^2k_x^2/2m^*_B$ situations.
\section{\label{sec:level1} Summary\protect\\ \lowercase{}}

We described spin-conserving and spin-flip opaque reflections of 2D electrons from a barrier in InSb/InAlSb heterostructure. In such a system the electric field of the barrier is the only source of spin splitting in the presence of spin-orbit interaction. Formulas for the  Bychkov-Rashba spin splitting are calculated for unbound electrons. Angles and amplitudes are calculated for reflected and penetrating electron beams. Two unexpected possibilities are predicted due to the spin-orbit interaction: a) non-attenuated electron propagation in the barrier when the latter is higher than the kinetic energy of incoming electrons, b) total reflection of electrons whose energies exceed barrier's height. It is shown that the considered system can serve as a source or filter of spin-polarized electron beams.
\ \\

{\bf Acknowledgments}
\ \\
We are grateful to Dr T. M. Rusin for elucidating discussions.
\ \\

\end{document}